\documentclass[11pt,twoside]{article}
\pdfoutput=1
\usepackage{asp2014}

\aspSuppressVolSlug
\resetcounters

\begin{document}

\title{Software solutions for numerical modeling of wide-field telescopes}

\author{
S.~Savarese$^1$, P.~Schipani$^1$, G.~Capasso$^1$, M.~Colapietro$^1$, S.~D'Orsi$^1$, M.~Iuzzolino$^2$, L.~Marty$^1$, F.~Perrotta$^1$, and G.~Basile$^1$
\affil{
$^1$INAF Osservatorio Astronomico di Capodimonte, Salita Moiariello 16, Naples, Italy\\
$^2$Officina Stellare S.r.l., Via Della Tecnica, 87/89, I-36030 Sarcedo (VI), Italy
}
\email{salvatore.savarese@inaf.it}
}

\paperauthor{Salvatore~Savarese}{salvatore.savarese@inaf.it}{https://orcid.org/0000-0002-4778-6050}{INAF Osservatorio Astronomico di Capodimonte}{}{Naples}{}{80131}{Italy}
\paperauthor{Pietro~Schipani}{pietro.schipani@inaf.it}{}{INAF Osservatorio Astronomico di Capodimonte}{}{Naples}{}{80131}{Italy}
\paperauthor{Giulio~Capasso}{giulio.capasso@inaf.it}{}{INAF Osservatorio Astronomico di Capodimonte}{}{Naples}{}{80131}{Italy}
\paperauthor{Mirko~Colapietro}{mirko.colapietro@inaf.it}{}{INAF Osservatorio Astronomico di Capodimonte}{}{Naples}{}{80131}{Italy}
\paperauthor{Sergio~D'Orsi}{sergio.dorsi@inaf.it}{}{INAF Osservatorio Astronomico di Capodimonte}{}{Naples}{}{80131}{Italy}
\paperauthor{Marcella~Iuzzolino}{}{}{Officina Stellare s.r.l.}{}{Sarcedo (VI)}{}{36030}{Italy}
\paperauthor{Laurent~Marty}{laurent.marty@inaf.it}{}{INAF Osservatorio Astronomico di Capodimonte}{}{Naples}{}{80131}{Italy}
\paperauthor{Francesco~Perrotta}{francesco.perrotta@inaf.it}{}{INAF Osservatorio Astronomico di Capodimonte}{}{Naples}{}{80131}{Italy}
\paperauthor{Giacomo~Basile}{giacomo.basile@inaf.it}{}{INAF Osservatorio Astronomico di Capodimonte}{}{Naples}{}{80131}{Italy}

  
\begin{abstract}

This paper presents an integrated modeling software to analyze the PSF of wide-field telescopes affected by misalignments. Even relatively small misalignments in the optical system of a telescope can significantly deteriorate the image quality by introducing large aberrations. In particular, wide-field telescopes are critically affected by these errors, insomuch that usually a closed-loop active optics system is adopted for a continuous correction, rather than for sporadic alignment procedures. Typically, a ray-tracing software such as Zemax OpticStudio is employed to accurately analyze the system during the optical design. However, an analytical model of the optical system is preferable when the PSF of the telescope must be reconstructed quickly for algorithmic purposes. Here the analytical model is derived through a hybrid approach and developed in a custom software package, designed to be general and flexible in order to be tailored to different optical configurations. First, leveraging on the Zemax OpticStudio API, the ray-tracing software is integrated into a Matlab pipeline. This allows to perform a statistical analysis by automatically simulating the system response in a variety of misaligned working conditions. Then, the resulting dataset is employed to populate a database of parameters describing the model.
  
\end{abstract}

\section{Introduction}
Misalignments of the optical elements in a telescope introduce non negligible aberrations which can significantly lower the image quality. In particular, wide-field telescopes \citep{capaccioli2011vlt,thomas2020vera} are so sensitive to these errors that a continuous correction by a closed-loop active optics system is normally necessary to get satisfactory performance.

The presence of misalignments does not yield additional aberrations in the overall optical system but rather a different combination of the existing ones, which can be modeled by means of the appropriate frameworks of the Nodal Aberration Theory \citep{shack1980influence,thompson1980aberration}.

During the optical design, the impact of misalignments on the performance is studied numerically, resorting to dedicated ray-tracing software. An accurate and widespread software is represented by Zemax OpticStudio. This kind of analysis is carried out in the early stages to give designers a sense of the mechanical tolerances and sensitivity of the telescope.

In applications where the PSF must be calculated many times, e.g. for iterative algorithmic procedures, an analytical model of the optical system is preferable to ray-tracing simulations which, although perfectly accurate, require a significant computational burden leading to unacceptable execution times.

The main focus of this paper is to present a general software solution to build an approximate analytical model of a wide-field telescope, capable of quickly estimating its wavefront error and PSF given the misalignment state of its optical elements. This is achieved thanks to an integrated modeling approach, leveraging a tight interaction between the optical simulations provided by Zemax and a custom processing pipeline implemented in Matlab.

\section{Model}
The main idea behind the model, just touched here for the sake of brevity, but more detailed in \citep{schipani2021modeling}, is to consider the Zernike expansion of the wavefront error $W$ in the pupil plane $\left(\rho,\vartheta\right)$:

\begin{equation}
    W\left(H_x,H_y,\rho,\vartheta,\zeta\right)=\sum_i {C_i}\left(H_x,H_y,\zeta\right){Z_i}\left(\rho,\vartheta\right)
    \label{eq1}
\end{equation}
where $\left(H_x,H_y\right)$ represent the image plane coordinates and $\zeta$ the perturbation vector.\\
According to this model, the expansion coefficients $C_i$ depend on the perturbation state representing a given misalignment, and on the field. This last dependence has been modeled through a Double Zernike (DZ) expansion.

\section{Software Implementation}

\articlefigure[width=0.6\textwidth]{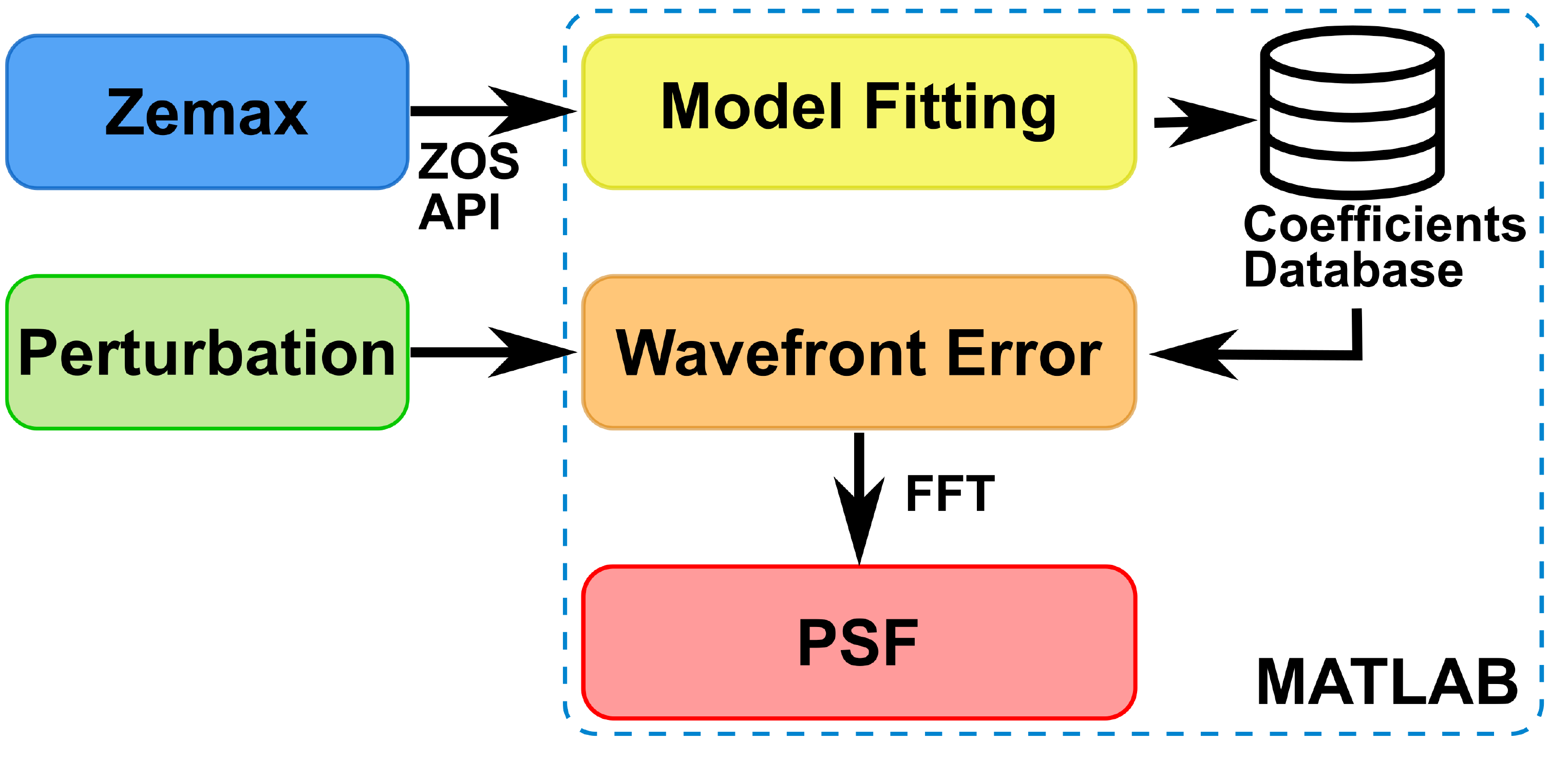}{fig1}{Integrated Modeling Software scheme.}

The wavefront analysis according to the adopted model is organized in two main steps, namely the model fitting for the coefficients database population and the wavefront analysis, which are detailed in the following sections. The logical scheme in Figure \ref{fig1} illustrates the software workflow.

\subsection{Model Fitting}

The first step of the elaboration concerns the evaluation of the model coefficients. To this end, an iterative procedure is run to simulate different misalignment conditions, in order to properly represent the perturbation space. For each one, Zemax is invoked to obtain the Zernike expansion of the wavefront error. This in turns allows to fit the particular coefficient $C_i$ according to the Double Zernike expansion.

Once established an analytic model, for a given telescope design this procedure may be executed offline and ideally only once to build a database of coefficients which can be then quickly queried during the following analysis. Nonetheless, given the great number of perturbations necessary to properly fit the model, performing this step manually using Zemax is cumbersome, and it is highly preferable to resort to the interface represented by the ZOS-API (Zemax OpticStudio Application Programming Interface).

\subsection{Wavefront Analysis and PSF reconstruction}

The Zemax simulation is run and governed by the main application developed in Matlab, which hosts the database and carries out the model computation. The resulting set of expansion coefficients allows various kind of analyses.

The first and most immediate is the reconstruction of the wavefront error according to Equation \ref{eq1}, from which the PSF can be readily obtained using the FFT method \citep{goodman2005introduction}.

\articlefigure[width=1\textwidth]{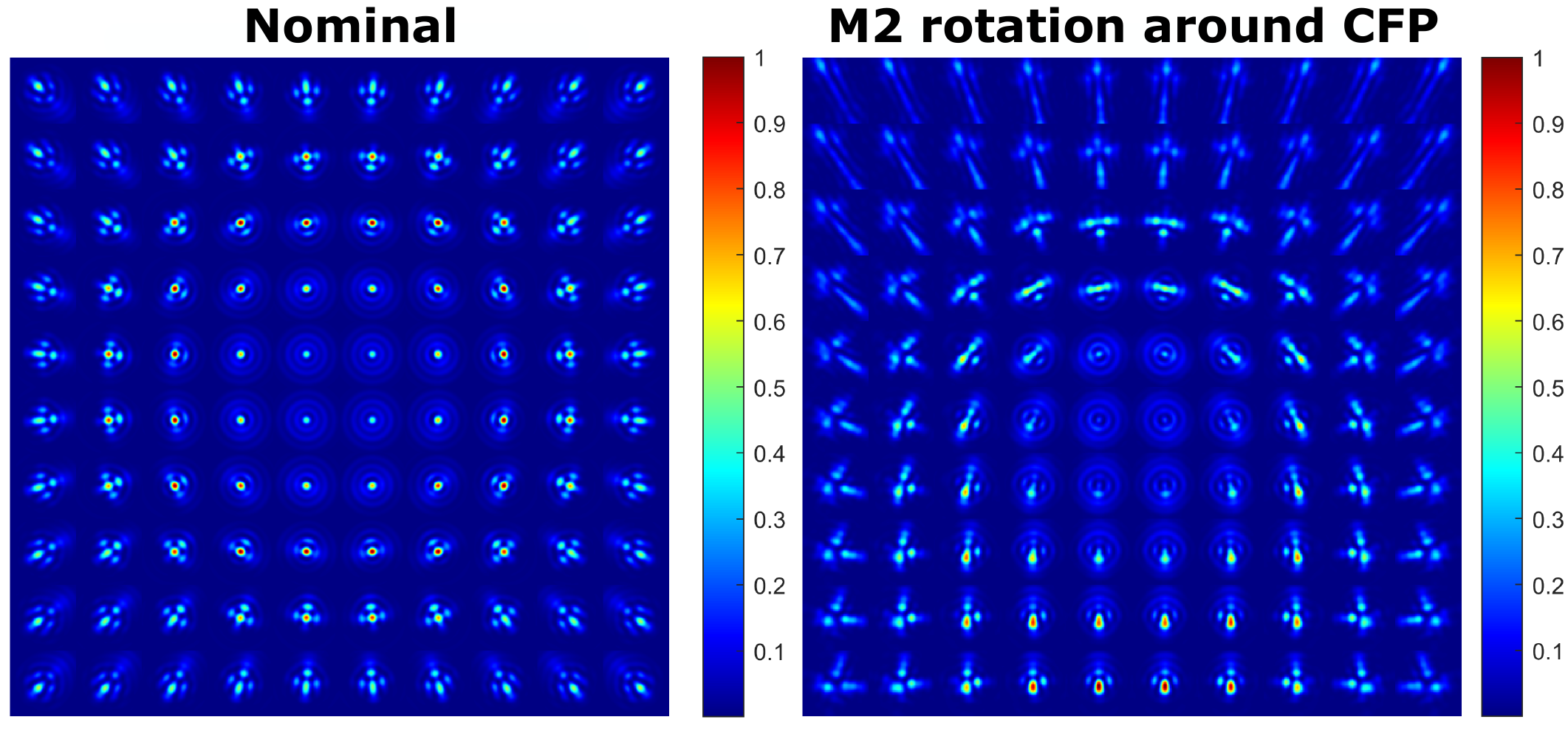}{fig2}{PSF map across the VST image plane in nominal conditions (left) and rotating the secondary mirror around the Coma Free Point (right).}
\articlefigure[width=1\textwidth]{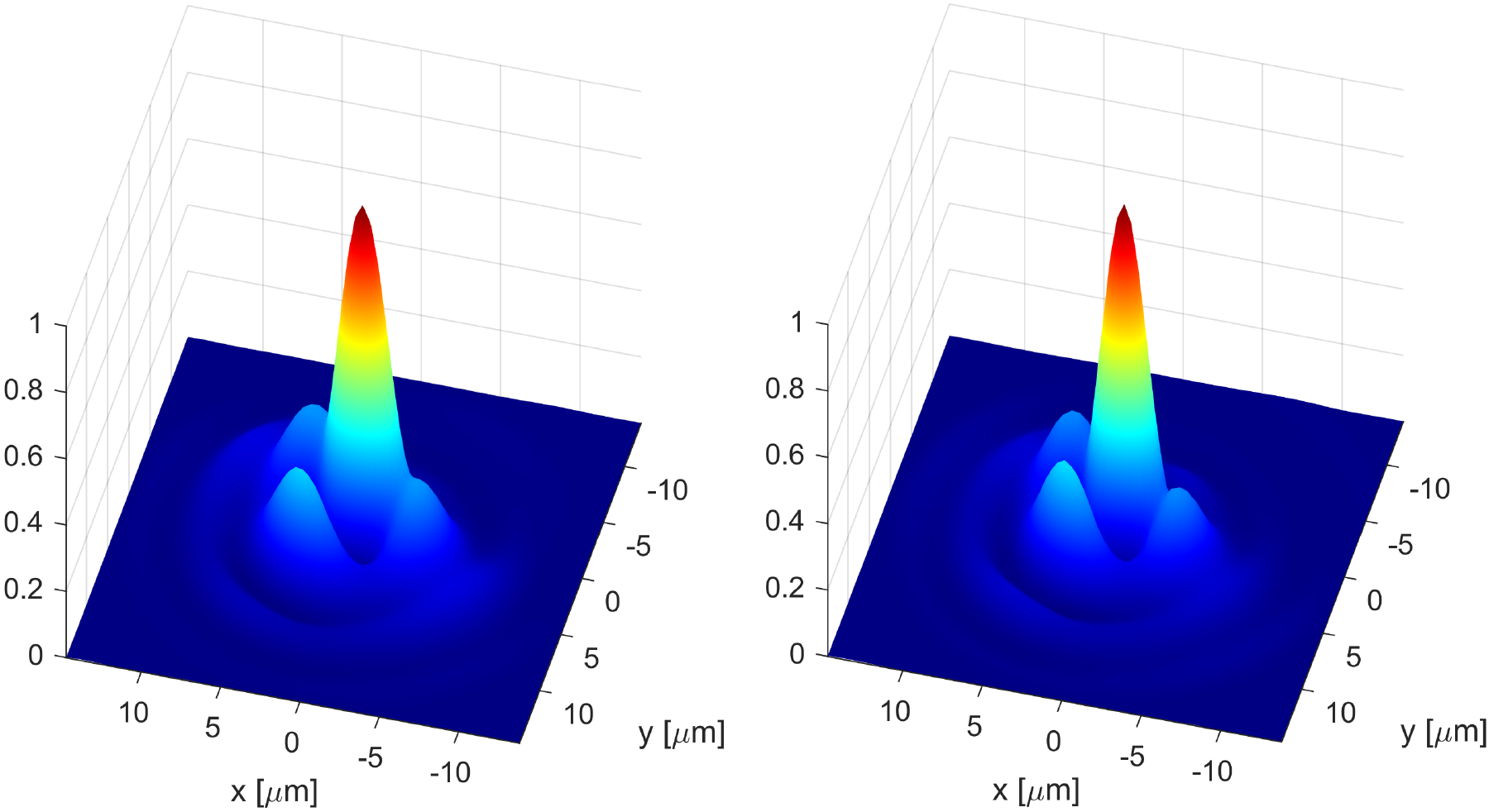}{fig3}{PSF Comparison between Zemax (left) and the analytic model (right).}

\section{Results}

As a case study, the model reconstruction is applied to the VST Telescope, characterized by a two-mirror modified Ritchey-Cr{\'e}tien optical configuration with a refractive field corrector.

First considering a null perturbation, the PSF for the telescope nominal (i.e. free from misalignments) condition has been computed on a matrix of points across the image plane, to produce the map shown in the left plot of Figure \ref{fig2}. Applying a rotation of the secondary mirror around the Coma Free Point, the corresponding aberrated PSF map is obtained, shown in the right plot of Figure \ref{fig2}.

The good accuracy of the software can be appreciated in Figure \ref{fig3}, where a single PSF computed with the analytical model is compared with the same one obtained by Zemax.

\section{Conclusions}

A software package has been developed to model the optical performance of wide-field telescopes in presence of misalignments.
The software has been implemented in Matlab and features a flexible external interface to Zemax OpticStudio to accurately simulate the telescope optical model. The simulation results are employed to build an analytic model able to quickly compute the wavefront error and the PSF for any perturbation state.
The PSF maps of the image field of the VST telescope under nominal and perturbed conditions have been presented as a test case, as well as a comparison between the analytical model and Zemax modeling of the PSF.

\bibliography{X2-003}

\end{document}